# YES, ABORIGINAL AUSTRALIANS CAN AND DID DISCOVER THE VARIABILITY OF BETELGEUSE

**Bradley E. Schaefer**
*Department of Physics and Astronomy, Louisiana State University,
Baton Rouge, Louisiana, 70803, USA*
Email: schaefer@lsu.edu

**Abstract**: Recently, a widely publicized claim has been made that the Aboriginal Australians discovered the variability of the red star Betelgeuse in the modern Orion, plus the variability of two other prominent red stars: Aldebaran and Antares. This result has excited the usual healthy skepticism, with questions about whether any untrained peoples *can* discover the variability and whether such a discovery is likely to be placed into lore and transmitted for long periods of time. Here, I am offering an independent evaluation, based on broad experience with naked-eye sky viewing and astro-history. I find that it is easy for inexperienced observers to detect the variability of Betelgeuse over its range in brightness from V = 0.0 to V = 1.3, for example in noticing from season-to-season that the star varies from significantly brighter than Procyon to being greatly fainter than Procyon. Further, indigenous peoples in the Southern Hemisphere inevitably kept watch on the prominent red star, so it is inevitable that the variability of Betelgeuse was discovered many times over during the last 65 millennia. The processes of placing this discovery into a cultural context (in this case, put into morality stories) and the faithful transmission for many millennia is confidently known for the Aboriginal Australians in particular. So this shows that the whole claim for a changing Betelgeuse in the Aboriginal Australian lore is both plausible and likely. Given that the discovery and transmission is easily *possible*, the real proof is that the Aboriginal lore gives an unambiguous statement that these stars do indeed vary in brightness, as collected by many ethnographers over a century ago from many Aboriginal groups. So I strongly conclude that the Aboriginal Australians *could* and *did* discover the variability of Betelgeuse, Aldebaran, and Antares.

## 1 INTRODUCTION

Betelgeuse ($\alpha$ Ori) is the bright red supergiant star in the shoulder of Orion, and it varies in brightness from visual magnitude 0.0 to 1.3 mag with a quasi-periodicity of 423 days. This variability was first 'discovered' by Sir John Herschel in 1836, or at least this was the first surviving observation to make it into Western science journals. Recently, there has been substantial excitement in the press and in the international variable star communities over a claim that the Aboriginal Australians had long ago discovered the variability of Betelgeuse (plus Aldebaran and Antares) and incorporated this discovery into lore passed down through untold generations, and being even today recognizable by ethnographers who have collected this lore.

The Aboriginal Australian discovery was first noted by Fredrick (2008) then described in more depth in Leaman and Hamacher (2014) and further analyzed in Hamacher (2018). They point to lore collected a century ago from southern and central Australia for multiple groups, with two basic stories plus the usual variants. The first story is about a hunter (represented by the modern stars of Orion) chasing after some young sisters (represented by the Pleiades), with his 'fire lust' (represented in the star Betelgeuse) waxing and waning, while the older protective sister's left foot (represented by Aldebaran) also fills and empties of the 'fire magic'. The second story is also a morality tale, with a story about a young male initiate (named Waiyungari) covered in red ochre running away into the sky, where he now sits in a canoe on the Milky Way, flanked by two stars representing two women, being prominent in the September evening skies. Waiyungari is definitively identified as Antares. The lore states that when Waiyungari brightens and gets hotter, this increases the sexual desire of the people. These are clear statements from multiple communities and multiple ethnographers that the three red stars get brighter and fainter.

This basic claim (that the Aboriginal Australians discovered the variability of Betelgeuse) has excited skepticism in some quarters. The unstated root of this skepticism is the view that Aboriginal people were too 'primitive' to be able to make a discovery that is more typical of modern Western science, as well as a general frustration from having heard so many implausible claims of discoveries of lost-ancient-wisdom. This skepticism can be itemized under the question "How could the Aboriginal Australians discover the variability when so many great Western astronomical observers from before 1836 all missed it?" and under the statement "The variability of Betelgeuse is too subtle and infrequent for any casual discovery".

In this note, I will be evaluating the basic claim that the Aboriginal Australians did discover the variability of Betelgeuse. This is just an independent and critical examination of prominent claims in the field of the history and heritage of astronomy, as our field so desperately needs.

## 2 *CAN* VARIABILITY BE DISCOVERED?

The first question is as to whether the variability of Betelgeuse *can* be detected by inexperienced observers with no variable star training?

Betelgeuse varies between visual magnitudes of 0.0 and 1.3. This can be directly visually compared with Capella (V = 0.08), Rigel (V = 0.13), Procyon (V = 0.38), Pollux (V = 1.14), Adhara (V = 1.50) and Bellatrix (V = 1.64). So it variously is brighter than nearby Capella and Rigel (and being substantially brighter than nearby Procyon), all the way down to being fainter than Pollux. These are blatant changes, made obvious by the good nearby comparison stars.

The human eye is good for comparing nearby stars, even with a need to glance back and forth (as well known to the many amateur variable star observers). Inexperienced people can detect substantial differences in brightnesses of near 0.3 mag, while well-experienced observers can get as good as 0.10 mag. These approximate numbers are well known, and I have confirmed them with a variety of experiments with many subjects, as well as large-scale experiments with the American Association of Variable Star Observers (AAVSO). What is needed is a small set of relatively nearby comparison stars, best with their magnitudes at the top, middle, and bottom of the range. And that is exactly what we have for Betelgeuse (and Aldebaran and Antares). Direct visual comparison of stars is easy for separations of ~60° or less, and indeed variable star observers have long made visual estimates while looking in eyepieces with apparent fields diameters of 60°. Rigel is 18° away, so this apparent separation is very easy for accurate and natural comparisons. Procyon, Pollux, and Capella are within 23° to 39°, all at comfortable distances for comparison. The primary visual comparison might be with the relatively nearby and connected star Procyon, and it will be easy to remember that Betelgeuse is brighter-than, roughly-equal-to, fainter-than, or much-fainter-than Rigel. With Betelgeuse's historical range of 0.0 to 1.3, its variability can be easily detected for either experienced or inexperienced observers.

Hamacher (2018) furthers claim that Australian Aboriginal lore contains unmistakable reports of the long-term pulsations of the red stars Aldebaran and Antares. Aldebaran is a red giant star that varies from 0.7 to 1.2 mag (from the AAVSO data) with variability time scales of around a year. Aldebaran has the same good comparison stars as Betelgeuse. Antares is a red supergiant star that varies from 0.7 to 1.7 mag (from the AAVSO data) with a pulsational timescale of 5.97 years on average. Antares has good comparison stars including Hadar (V = 0.63), Altair (V = 0.93), Spica (V = 1.05), Acrux (V = 1.27), Mimosa (V = 1.30) and Shaula (V = 1.62). With these amplitudes, time scales, and comparison stars, the variability of both Aldebaran and Antares can be easily detected for either experienced or inexperienced observers.[1]

The changes are so blatant that written records are not needed, as it is easy to remember that Betelgeuse is, say, greatly brighter than Procyon, whereas in the past year it was greatly fainter than Rigel. That is the sort of comparison that humans are good at.

So my strong conclusion is "Yes, the ordinary variability of Betelgeuse is easily discovered by inexperienced observers."

## 3 *COULD* THE DISCOVERY BE PASSED ON IN LORE?

As an astronomy historian, I specialize in visual photometry in old times (see for example, Schaefer, 1993; 1996; 2001; 2013a; 2013b), and I recognize that it takes more than just ability for a culture to make a discovery. In the case of John Herschel, he was explicitly measuring magnitudes of all stars in the sky, and this is how he discovered Betelgeuse's variability, whereupon he passed his finding into the scientific literature. Pliny tells us that variability-searches are the exact reason why Hipparchus created his original star catalog. But why should Aboriginal Australians (or any other culture outside of the Greek-descended tradition) be looking for variability, much less pass it on?

Well, there is no particular reason to expect that the Aboriginal Australians did any systematic variability-searching. But systematic searches are not needed for discovery. For the many waking hours during each night, the only light away from the campfire is up in the stars, so these were frequently seen throughout the lives of Aboriginal people. Betelgeuse is among the brightest dozen or so stars, while it is in a striking pattern in modern Orion, and has a striking color—so if you are repeatedly looking at stars, then Betelgeuse would be a natural and frequent target. This always-under-the-skies lifestyle for Aboriginal people with the dark Outback skies is completely alien to modern scholars who rarely venture outside of our modern light-polluted skies.

Indigenous people the world over are renown for exhaustive observations of all of nature around them (e.g. Norris, 2016; Ruggles, 2011), as it is critical for survival. So it is inevitable that many Aboriginal people over the millennia would watch Betelgeuse sufficiently closely to easily recognize that it varies from brighter-than to much-fainter-than Procyon. No written records are needed for such observations, and the memory of prior brightness extremes needs to be kept for only half a year or so.[2] Thus, it is reasonable and inevitable that many Aboriginal people (and many people from other cultures) over the millennia discovered the variability of Betelgeuse.

A possible counter-argument is that Betelgeuse's variability was missed by excellent Western observers, such as Ptolemy, Tycho Brahe, and William Herschel, and hence was not discovered by the Aboriginal Australians. The logic of this counter-argument is unusably poor because every sky phenomenon has a first surviving discoverer, but this says nothing about earlier observers whose discoveries might not be documented in refereed journals. Arguments by historic ignorance are poor logic. One pervasive and well-documented case is that the existence of atmospheric extinction was never reported anywhere in any source in any way before 1727, yet Ptolemy, Al-Sufi, Tycho, and Hevelius all independently discovered the dimming of light by our atmosphere. They modeled the quantitative dimming in magnitudes as a function of altitude, and reported extinction-corrected magnitudes in their star catalogs (Schaefer 2013b), yet there is no record of their measurements or even that they knew the atmosphere dims light. And are we really to say that the first record in science journals of the Green Flash invalidates Jules Verne's earlier book highlighting the Green Flash (Le Rayon Verde) dated 1882, or indeed that myriads of sunset watchers the world over have not made many independent discoveries of the Green Flash? As another example, the phenomenon of mountain-shadows-cast-onto-air-as-cones-of-darkness was first published in the early 1900s, yet this does not invalidate the 1789 painting from the top of Snaefellsjokull that clearly shows the phenomenon (Pocock, 1791), nor does it invalidate the correct description of the shadow from the Canary Islands in Arthur Conan Doyle's 1884 "J. Habakuk Jephson's Statement", nor does it invalidate the long lore of mountain shadows for Mount Fuji and for Mauna Kea. And so on. The point is that just because one-good-person misses a phenomenon, this means nothing about whether the Aboriginal Australians, or Sioux Indians, or Victorian-era writers independently discovered the same phenomenon.

Another question is whether a discovery gets saved as a report that survives to be seen by modern scholars. Certainly, a large fraction of discoveries are lost to posterity for any of many reasons. In literate societies, the discovery must be written down and that record somehow saved until modern times. In cultures without writing, the discovery must become part of the lore passed along through the generations, and continue on long enough to be recorded by an ethnographer. Nevertheless, there are well known paths for putting the discovery into recognizable lore, and for reliably passing the lore on through many centuries.

For putting the discovery into lore, it must be made somehow culturally relevant. Let me give some astronomical examples:

(1) The 8th-millenium tightest planetary massing on 26 February 1953 BCE was made culturally relevant for the ancient Chinese as being the start date for calendars and as inspiring the start of the *Mandate of Heaven* doctrine (Schaefer, 2000). This was passed along orally for over a millennium before being written down.

(2) For naked-eye variable stars, the variations of Algol passed into the calendars of ancient Egypt (Jetsu et al., 2013) and passed into complex mythology associated with Medusa, as appropriate for the sky, by the ancient Greeks (Wilk, 2000).

(3) The knowledge of the very rare total solar eclipses (every 310 years on average from any one location) has passed into the lore of all old cultures worldwide as being one of the worst omens possible (Schaefer, 1994). The universal abject fear of these events can only be remembered by the events entering as lore passed down for many generations. This conversion of a celestial event into lore is applicable in Australia (Norris, 2016: Section 2.4).

(4) Many more Aboriginal Australian examples are cited in the review of Norris (2016), including the Emu in the sky (mostly defined by the dark clouds in the Milky Way from the Coalsack to Sagittarius) showing a depiction of the flying bird with an explanation for how the emu lost the ability to fly.

The point of these examples is to show how sky events/pictures can get passed into the traditions and stories of indigenous cultures for a wide variety of reasons. In particular, the lore involving the three red stars was incorporated into Aboriginal traditions as morality tales.

Once in an oral culture's lore, the information must pass through many generations without substantial garbling. Famous examples are the epic poetry of Homer passed for many centuries, the chief lists and navigation records of the Polynesians passed for a millennium, and sky lore of the Great Bear passed along after crossing the Bering Strait for over 14 millennia. The theory of this faithful transmission is well described (Kelly, 2015). Aboriginal Australian examples include coastline flooding roughly 7 millennia ago (Nunn and Reid, 2016), reports of the meteorite fall that formed the Henbury craters approximately 4 millennia ago (Hamacher, 2013), and the Great Eruption of the star $\eta$ Carinae about 150 years ago (Hamacher and Frew, 2010). So we can be sure that lore is often faithfully transmitted for many centuries and millennia.

## 4 *DID* ABORIGINAL AUSTRALIANS DISCOVER BETELGEUSE'S VARIABILITY?

So the variability of Betelgeuse is readily discovered by Aboriginal Australians. Indeed, it is inevitable that Aboriginal people would make this discovery many times over their many millennia of close nature observing under the wonderful Outback skies, and such a discovery can easily be placed into Aboriginal lore, and then passed down many generations until recorded by ethnographers and anthropologists interviewing elders. So the whole claim of Aboriginal Australian discovery of Betelgeuse's variability is easily possible. But many ideas for transmissions of discoveries are *possible*. So the real question is not whether the Aboriginal Australians *could* make the discovery and pass this along as lore, but whether they *did*.

I have spent about 25% of my career in astro-history and celestial visibility (specializing in visual photometry) debunking ideas where someone claims that some astronomical event is possible and hence such-and-so culture *must* have observed it (see for example Schaefer, 1983; 1991; 2006). These claims and reports all have the unifying and critical issue that they offer zero positive evidence other than the one lore or text item that they are trying to explain. Often these claims are fairly spectacular, represent an isolated report, are often anachronisms that are not plausible within the cultural context, and typically run afoul with technical errors. The variability of Betelgeuse is not spectacular. It is reported by multiple independent communities, for other known red variable stars easily fits into the cultural context, and has no technical problems.

My knee-jerk reaction is to be skeptical of any claim for ancient knowledge, and many others in our community might be similarly biased. The reason is that we see so many bad claims, usually for exotic discoveries, where there is no positive evidence. Wrong biases against the Betelgeuse-variability-discovery result also arise in other communities. For example, the astro-history community has a bias against anything that does not have documenttation. Under this bias, the ethnographic reports are not counted as 'documentation'. Other communities believe that Indigenous peoples were incapable of making any discoveries paralleling something from Western science. Such argumentation is wrong. The existence of bad claims with no evidence has no relevance for a claim supported by significant positive evidence and excellent ethnographic documentation from oral cultures with complex knowledge of nature. I am stating these various biases explicitly, so that we all can hopefully expunge them from our inner thinking.

Another reason to often be skeptical of such claims for ancient knowledge is that the lore or text is ambiguous, requiring some arcane decoding. Yes, old lore and documents can and do report events and facts in symbolic or poetic language, but there are a myriad of possible decodings for anything other than blatant symbolism. So in general, decoding is fruitless because any interpretation is just indistinguishable as one of many. But we do not have such ambiguities for the cases of the three red stars. There is adequate sky lore describing the constellations such that the identification of the stars is clear, with no plausible alternatives. And the Aboriginal lore explicitly says that the star changes in brightness, with recurring waxing and waning.

For the question of whether the Aboriginal Australians *did* discover the variability of the three red stars, the direct words and ethnographic reports provides the proof. For Betelgeuse and Aldebaran, we have firsthand ethnographic documentation from Daisy Bates in 1921 from Aboriginal groups in the Great Victoria Desert. For Antares, we have ethnographic documentation from South Australia by H.A.E. Meyer in 1846, Norman Tindale in 1935, and Philip Clarke in 1999. Aboriginal people *do* tell us, many times over, from multiple communities, that these stars vary in brightness. The ethnography provides the *proof* that Aboriginal people actually *did* discover the variability of Betelgeuse.

## 5 CONCLUSION

So the case for the Aboriginal Australian discovery of the variability of Betelgeuse, Aldebaran, and Antares, is:

(1) It is easily possible to detect variability from 0.0 mag to 1.3 mag by inexperienced observers;
(2) We know that the Aboriginal people were intensive and exhaustive observers of their natural environment, including the stars, so Betelgeuse, as one of the brightest dozen stars and distinctly colored, will have been watched many times over by many Aboriginal people over the past 50 millennia;
(3) Some fraction of these many watchers will remember recent times when Betelgeuse was at an obviously different state, hence discovering its variability;
(4) Only some small fraction of these discoveries need be saved into the lore, with the

reason being that it was placed into morality stories;

(5) Once into the Aboriginal lore, we know that this information can be transmitted reliably over centuries and millennia.

(6) But the real proof for the Aboriginal discovery of these stars' variability is that the Aboriginal people themselves report that they vary in brightness.

This review of the claim for Aboriginal discovery of Betelgeuse's variability is also a review of the general methodology and results of pulling astronomical information from the ethnographic record, both for Aboriginal Australians in particular and many ancient or oral cultures in general. What I have found is partly a mixed set of evaluations. Many claims coming from fringe workers have no compelling evidence, are contrary to cultural plausibility, and are often technically wrong. Nevertheless, many claims are confident with plenty of evidence, cultural plausibility, and no technical problems. What I am concluding is that the basic methods are fine and convincing, but care must be taken to recognize and marginalize fringe claims. In particular, fringe claims tend to dominate the press and the internet, so inexperienced people can easily be misled into accepting claims that are known-wrong. This can skew and screw the whole field, with archaeoastronomy being an example, where the majority of press reports and popular knowledge are known to be flat wrong. It is unclear how to handle the flood of evocative fringe claims, but our scholarly community must start by making evaluations of claims, and not shy away from declaring 'Wrong!'.

For the field of Australian Aboriginal astronomy with attention to variable stars, we also see the full range of claims made. We see evidence-less claims for supernova depictions in Aboriginal rock art (Murdin, 1981), with this just being part of a worldwide movement to depict any old squiggle as a supernova. But we also see Hamacher (2014) correctly rejecting all of these erroneous claims, and establishing a set of criteria for confirming novae/supernovae descriptions in oral traditions or motifs in material culture. Also, Hamacher and Frew (2010) demonstrated that the weird and unique eruption of the Luminous Blue Variable star η Carinae in the 1840s was transferred into the oral traditions of the Boorong peoples in Victoria, near Lake Tyrrell, and proved this connection with concurrent ethnographic reports. Now, Leaman and Hamacher (2014) and Hamacher (2018) report Aboriginal discovery of the variability of three red giant stars, with this discovery being both easy and inevitable, while we know that the discoveries have been placed into lore multiple times, as proven from the ethnographic record. The last three examples provide good exemplars for future workers in our field.

In summary, we can be very confident that Aboriginal Australians did discover the variability of Betelgeuse, Aldebaran and Antares, and described them in oral traditions that were passed down to modern times.

## 6  NOTES

1. We need not be worried by effects such as differential extinction or the Pukinje effect. For Australian skies, the stars of Orion always reach high altitudes with typical extinction coefficients of 0.3 mag/airmass, so we have negligible differential extinction. The amateur astronomer's Purkinje Effect, with red stars changing in apparent brightness as the star is stared at, only affects stars seen with night vision, whereas Betelgeuse is so bright as to be well above the threshold between day/night vision (V ≈ 3.1), as proven by Betelgeuse appearing reddish in color.

2. Betelgeuse cycles between maxima and minima with a typical time scale of half the pulsation quasi-periodicity of 423 days.

Dr Bradley E. Schaefer is the Distinguished Professor in the Department of Physics and Astronomy at the Louisiana State University in Baton Rouge, Louisiana, USA. He is a winner of the 2007 Gruber Prize for Cosmology and the 2015 Breakthrough Prize in Fundamental Physics, as one of the discoverers of the still-enigmatic Dark Energy and the accelerating Universe. With 230 publications in refereed journals, his astrophysics specialty has been variable stars of many types, concentrating on supernovae and novae.

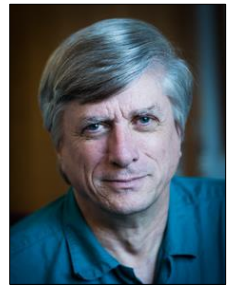

Seventy of these publications are on the history of astronomy, with a unique specialization in celestial visibility, in what constitutes a separate career. But really, Schaefer is just an avid amateur astronomer, even now running many visual projects and sky-tourism. For example, he has been a member of the American Association of Variable Star Observers (AAVSO) since 1972, working on a wide variety of programs. Starting in May 2018, Schaefer will travel to Australia, departing in November 2018, with part of the purpose being to do visual photometry and sight-seeing in the great Outback skies. One of the programs for Australia is to (for the second time) completely map the entire sky for all stars as viewed with the unaided eye.